\def\bibtex{0}  
\newcommand{\ifbibtex}{\ifnum\bibtex>0}
\documentclass[12pt]{iopart}

\usepackage{amssymb}
\usepackage{comment}
\usepackage{stackrel}
\usepackage{graphicx}
\newcommand{\bm}[1]{\mbox{\boldmath$#1$}}
\newcommand{\bra}[1]{\langle#1|}
\newcommand{\ket}[1]{|#1\rangle}

\newcommand{\mean}[1]{\langle#1\rangle}
\newcommand{\braket}[2]{\langle#1|#2\rangle}

\newcommand{\ketbra}[2]{|#1\rangle\langle#2|}

\renewcommand{\aa}{\hat a}
\newcommand{\ad}{\hat a^\dag}

\newcommand{\hA}{\hat A}

\newcommand{\hB}{\hat B}

\newcommand{\hX}{\hat X}

\newcommand{\n}{\hat n}

\newcommand{\hD}{\hat D}
\newcommand{\hN}{\hat N}
\newcommand{\hM}{\hat M}
\newcommand{\hU}{\hat U}
\newcommand{\hE}{\hat E}

\newcommand{\s}{\hat s}

\newcommand{\hW}{\hat W}

\newcommand{\hPi}{\hat\Pi}

\newcommand{\R}{\mathrm {Re\,}}
\newcommand{\I}{\mathrm {Im\,}}

\newcommand{\hI}{{\hat I}}

\begin{document}

\title[Quantum measurement and uncertainty relations in photon polarization]
{
Quantum measurement and uncertainty relations in photon polarization
}

\author{Keiichi Edamatsu}

\address{Research Institute of Electrical Communication, Tohoku University, Sendai 980-8577, Japan}
\ead{eda@riec.tohoku.ac.jp}
\vspace{10pt}
\begin{indented}
\item[]April 25, 2016
\end{indented}

\begin{abstract}
Recent theoretical and experimental studies have given raise to new aspects 
in quantum measurements and error-disturbance uncertainty relations.
After a brief review of these issues, 
we present an experimental test of the error-disturbance uncertainty relations in photon polarization measurement.
Using generalized, strength-variable measurement of a single photon polarization state,
we experimentally evaluate the error and disturbance
in the measurement process
and demonstrate the validity of recently proposed uncertainty relations. 
\end{abstract}

%
\vspace{2pc}

\section{Introduction}
Measurement (or observation) is the most fundamental issue in science and technology.
We understand the laws of nature only through measurements.
In classical science, we tacitly assume the existence of perfect measurement, 
in which one can measure a physical observable, e.g., a particle's position, very accurately without disturbing the object's state. 
This property ensures the objectivity of the measurement outcome and thus the reality of physical observable.
However, in quantum mechanics, such perfect measurement does not exist; 
any measurement with finite accuracy inevitably disturbs the object's state.
Heisenberg considered such a situation in his famous thought experiment on the gamma-ray microscope \cite{Heisenberg:1927gz}.
Since then, problems on quantum measurement and the uncertainty relation have long been discussed in fundamental quantum physics.
However, surprisingly, no commonly agreed definitions have been established on the measurement error and disturbance, 
and thus on uncertainty relations in quantum measurements.
Recent theoretical progress in quantum measurements and uncertainty relations
has revealed the new aspects of these issues \cite{Ozawa03a,Ozawa:2014jy,Branciard:2013cb,Busch:2014dc,Buscemi:2014ew}.
In addition, the concept of weak measurement and weak value \cite{Aharonov:1988fk} has attracted great attention.  

Photons are the quanta of electromagnetic waves.
A photon propagating in a vacuum has two degrees of freedom for its polarization, 
corresponding to two orthogonal directions of its oscillatory electric fields perpendicular to the propagation direction. 
Thus the polarization state of a photon can be treated as a two-level qubit system,
one of the simplest and most fundamental systems in quantum physics.
This polarization qubit held in a single photon is extremely useful in quantum information and communication technologies.
 However, as mentioned above, 
quantum measurement of even the simplest system has not yet been fully understood.  
In this article, 
I describe our recent experiments in which we realize the generalized measurement of the photon polarization qubit 
and evaluate its uncertainty relations.

This article is organized as follows.
In Sec.~2, a quantum-mechanical view of photon polarization is presented, mainly for non-specialists.
In Sec.~3, the general theory of quantum measurement is briefly reviewed.
In Sec.~4, definitions of error and disturbance in quantum measurements are introduced and discussed. 
In Sec.~5, uncertainty relations in quantum measurement are introduced, 
by contrast with the uncertainty relations in quantum state preparation.
In Sec.~6, our experimental results on the generalized measurement of photon polarization are presented
and compared with the uncertainty relations.
Secion~7 is the summary.

\section{Quantum optics in photon polarization} 
\label{sec.QMPP}

Here we consider a plane electromagnetic wave propagating 
along the $z$ axis.
Then, we take  the $x$ and $y$ axes as the two orthogonal directions in the oscillatory field plane.
The field along each axis is expressed by a harmonic oscillator and thus
the fileds in the $x$-$y$ plane are expressed by the two-dimensional harmonic oscillator. 
Let $\ad_i$ and $\aa_i$ ($i=x$ or $y$) be the creation and annihilation operators of the field parallel to $x$ or $y$, respectively  
and $\n_i=\ad_i\aa_i$ the corresponding number operator. 
A simultaneous eigenstate of 
$\n_x$ and $\n_y$ can be expressed as
$ \ket{n, m} $,
where $n$ and $m$ are the eigenvalues of  $\n_x$ and $\n_y$, 
i.e., the photon numbers in the $x$ and $y$ polarization modes.
In general, a pure state $\ket{\psi}$ of the two-mode field can be expressed as
\begin{eqnarray}
 \ket{\psi} = \sum_{n,m} c_{n,m}\ket{n,m}.
\end{eqnarray}
In classical optics, the polarization state is often characterized by the Stokes parameters \cite{Born:1999un}.
In quantum optics, the Stokes parameters turn out to be a set of operators, i.e., 
the Stokes operators defined as
\begin{eqnarray}
\s_0 &= \ad_x \aa_x + \ad_y \aa_y  =\n_x +\n_y ,
\label{ho16-12-0} \\
\s_1 &= \ad_x \aa_x - \ad_y \aa_y  =\n_x -\n_y ,
\label{ho16-12-1} \\
\s_2 &= \ad_x \aa_y + \aa_x \ad_y  ,
\label{ho16-12-2} \\
\s_3 &= -i (\ad_x \aa_y - \aa_x \ad_y ) ,  
\label{ho16-12-3} 
\end{eqnarray}
where $\s_0$ corresponds to the total photon number, 
$\s_1$, $\s_2$ and $\s_3$ present the degree of polarization in  
$x$-$y$, $\pm 45^\circ$, and left and right circular polarizations, respectively \cite{StokesOperatorsNote}. 
The Stokes operators are often used to characterize the polarization state of the quantized optical field.

Hereafter, we consider {\it single photon polarization states}, in which a single photon stays in either of the polarization modes,
i.e., $n+m=1$. 
A pure state of the single photon polarization is expressed by the linear combination of 
$ \ket{1, 0} $ 
and 
$ \ket{0, 1} $:
\begin{eqnarray}
 \ket{\psi} 
 &= 
 \alpha \ket{1, 0}  +  \beta \ket{0, 1},
\label{qf6-7} 
\end{eqnarray}
where $|\alpha|^2+|\beta|^2=1$.
For this state, we find
\begin{eqnarray}
\mean{\aa_x} =\mean{\ad_x}= \mean{\aa_y} =\mean{\ad_y}= 0
\label{qf6-9} 
\end{eqnarray}
and thus the mean values of all the field amplitudes are zero. 
Nonetheless, the mean value of the Stokes operators are obtained as:
\begin{eqnarray}
\mean{\s_0} &=  |\alpha|^2 + |\beta|^2 = 1 ,\\
\mean{\s_1} &= |\alpha|^2 - |\beta|^2  ,\\
\mean{\s_2} &=  2 \R (\alpha^* \beta) ,\\
\mean{\s_3} &=  2 \I (\alpha^* \beta) .
\label{qf6-10} 
\end{eqnarray}
These are equivalent to those of the classical pure polarization state associated with the field amplitude vector
(Jones vector)
$ (\alpha, \beta) $. 
Thus the single photon polarization state (\ref{qf6-7}) is a pure polarization state
satisfying
\begin{eqnarray}
\mean{\s_1}^2 + \mean{\s_2}^2 + \mean{\s_3}^2 
= \mean{\s_0}^2 ,
\label{qf6-101} 
\end{eqnarray}
where
the Stokes vector 
$\left( \mean{\s_1}, \mean{\s_2}, \mean{\s_3} \right)  $
reaches the surface of the Poincar\'e sphere with the radius $\mean{\s_0}$.

The two bases, 
$ \ket{1, 0} $ 
and 
$ \ket{0, 1} $,
in (\ref{qf6-7})
are often written as 
\begin{eqnarray}
 \ket{1, 0} \equiv \ket{\mathrm H}, \quad  \ket{0, 1} \equiv \ket{\mathrm V},
\label{qf6-11} 
\end{eqnarray}
where H and V mean horizontal and vertical polarizations, respectively.
Then, the single photon polarization state can be expressed as
\begin{eqnarray}
 \ket{\psi} 
 &= 
 \alpha \ket{\mathrm H}  +  \beta \ket{\mathrm V}.
\label{qf6-12} 
\end{eqnarray}
It is also convenient to define  
the linear polarization states along $\pm45^\circ$ directions, 
$\ket{\mathrm D}$ and $\ket{\mathrm A}$,
as
\begin{eqnarray}
\ket{\mathrm D} 
\equiv \frac{1}{\sqrt{2}} \left( \ket{\mathrm H} + \ket{\mathrm V} \right), \quad
\ket{\mathrm A} 
\equiv \frac{1}{\sqrt{2}} \left( \ket{\mathrm H} - \ket{\mathrm V} \right),
\label{qf6-14} 
\end{eqnarray}
and 
the left and right circular polarization states,
$\ket{\mathrm L}$ and $\ket{\mathrm R}$,
as
\begin{eqnarray}
\ket{\mathrm L} 
\equiv \frac{1}{\sqrt{2}} \left( \ket{\mathrm H} + i \ket{\mathrm V} \right), \quad
\ket{\mathrm R} 
\equiv \frac{1}{\sqrt{2}} \left( \ket{\mathrm H} - i \ket{\mathrm V} \right).
\label{qf6-16} 
\end{eqnarray}
These 
are the eigenstates of $\s_1$, $\s_2$, and $\s_3$:
\begin{eqnarray}
  \s_1 \ket{\mathrm H} = \ket{\mathrm H}, &\quad   \s_1 \ket{\mathrm V} = -\ket{\mathrm V},
\label{qf6-13}  \\
  \s_2 \ket{\mathrm D} = \ket{\mathrm D}, &\quad   
  \s_2 \ket{\mathrm A} = -\ket{\mathrm A}, 
\label{qf6-15}  \\
  \s_3 \ket{\mathrm L} = \ket{\mathrm L}, &\quad   
  \s_3 \ket{\mathrm R} = -\ket{\mathrm R} .
\label{qf6-17}  
\end{eqnarray}
The single photon polarization state is expressed by these bases
in the two-dimensional Hilbert space; 
it is thus expressed by SU(2) algebra as in the spin $1/2$ system.
In practice,
using $\ket{\mathrm H}$ and $\ket{\mathrm V}$ as the bases,
the matrix representations of the Stokes operators
result in the Pauli matrices \cite{PauliMaticesNote}: 
\begin{eqnarray}
  \sigma_0=I=\left(\begin{array}{cc}1 & 0 \\0 & 1\end{array}\right), &\quad
  \sigma_1=\sigma_z=\left(\begin{array}{cc}1 & 0 \\0 & -1\end{array}\right), \nonumber \\
  \sigma_2=\sigma_x=\left(\begin{array}{cc}0 & 1 \\1 & 0\end{array}\right), &\quad 
  \sigma_3=\sigma_y=\left(\begin{array}{cc}0 & -i \\i & 0\end{array}\right) .
\label{eq.PauliOperators} 
\end{eqnarray}
Thus, 
where the single photon polarization states are concerned,
the Stokes operators are equivalent to the Pauli matrices. 
Accordingly, the Stokes vector 
and the Poincar\'e sphere 
for the single photon polarization states
are equivalent to the Bloch vector and the Bloch sphere for a two-level qubit system, respectively.

\begin{figure}[bt]
\centerline{
\includegraphics[width=90mm]{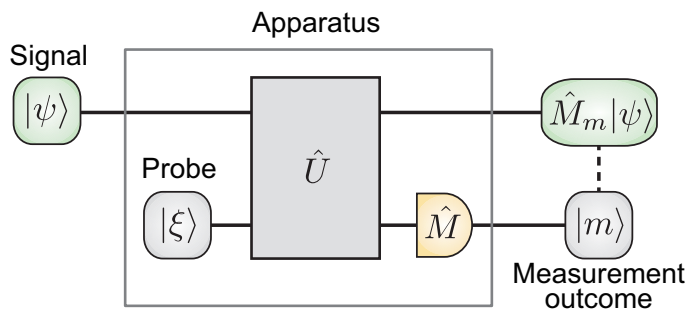}
}
\caption{
Model of indirect quantum measurement.
The signal state $\ket{\psi}$
and the probe (meter) state $\ket{\xi}$
interact with each other through the unitary operator $\hU$.
Because of $\hU$, the observable $\hA$ on the signal can be correlated with the observable $\hM$ on the probe.
Then $\hM$ is measured on the probe. 
This procedure indirectly measures $\hA$ on the signal.
}
\label{fig.model}
\end{figure}

\section{Quantum measurement}

\subsection{Projective measurement}
Here we consider a model of the quantum measurement of an observable  $\hA$
with a discrete spectrum and a finite dimension.   
The spectral decomposition of $\hA$ is expressed as
\begin{eqnarray}
\hA = \sum_{j} \lambda_j \ketbra{j}{j} = \sum_{j} \lambda_j \hPi_{j},
\label{eq.SD1}
\end{eqnarray}
where  $\lambda_j$ is the eigenvalue of $\hA$ and $\hPi_{j}\equiv\ketbra{j}{j}$ is the projector to the corresponding eigenstate $\ket{j}$.
The {\it projective measurement} of $\hA$ observes the state in one of the eigenstates $\ket{j}$, 
and assigns the outcome $\lambda_j$.
When the projective measurement acts on the state $\ket{\psi}$,
the probability $P(j)$ to find the state in $\ket{j}$ is given by 
\begin{eqnarray} 
P(j)=\braket{\psi}{j}\braket{j}{\psi} 
=\mean{\hPi_j}, 
\label{eq.probA}
\end{eqnarray}
where the avarage
$\mean{...} $ is taken over the signal state $\ket{\psi}$.
The mean value of the measurement of $\hA$ for the state $\ket{\psi}$ is obtained as
\begin{eqnarray}
\mean{\hA} 
=  \sum_{j} \lambda_j \mean{\hPi_j} .
\label{eq.meanA}
\end{eqnarray}

\subsection{Generalized measurement}
\label{sec.GM}
We introduce a model of a measurement instrument in which the signal $\ket{\psi}$ interacts with a probe (meter) state $\ket{\xi}$
through the interaction unitary operator $\hU$,
as shown in Fig.~\ref{fig.model}.
After the interaction, the initial state 
$ \ket{\Psi} = \ket{\psi}\otimes \ket{\xi}$
is converted to 
\begin{eqnarray}
\ket{\Psi} \stackrel[\hU]{}{\longrightarrow} \ket{\Psi'} = \hU \left( \ket{\psi}\otimes \ket{\xi} \right) .
\end{eqnarray}
Then, we make the projective measurement of observable $\hM$ on the probe state;
the spectral decomposition of $\hM$ is given by
\begin{eqnarray}
\hM = \sum_m \mu_m \ketbra{m}{m},
\end{eqnarray}
where  $\mu_m$ is the eigenvalue of $\hM$ and $\ket{m}$ the corresponding eigenvector.
The state $\ket{\Psi'}$ after the interaction can be decomposed 
in terms of $\ket{m}$:
\begin{eqnarray}
\ket{\Psi'} = \sum_m \left(\hM_m \ket{\psi}\right)\otimes\ket{m},
\label{eq.GM}
\end{eqnarray}
where $\hM_m$ is the {\it measurement operator} acting on the signal state $\ket{\psi}$:
\begin{eqnarray}
\hM_m = \bra{m}\hU \ket{\xi}.
\label{eq.GM2}
\end{eqnarray}
The probability $P(m)$ to find the probe state in $\ket{m}$ is
\begin{eqnarray}
P(m)
= \mean{\hM_m^\dag \hM_m }
= \mean{\hE_m }
\label{eq.POVM1}
\end{eqnarray}
where 
$\hE_m = \hM_m^\dag \hM_m $
is the {\it positive operator valued measure} (POVM) element satisfying
\begin{eqnarray}
\sum_m \hE_m = \hI ,
\label{eq.POVM3}
\end{eqnarray}
where $\hI$ is the identity operator.
Thus the POVM element $\hE_m$ 
determines the probability of finding the measurement outcome in $m$.
The signal state is changed from its initial state $\ket{\psi}$ by $\hM_m$;
the measurement operator characterizes the backaction of the measurement. 
The measurement system characterized by Eq.~(\ref{eq.GM}) is referred to as {\it generalized measurement}.
When $\hA$ and $\hM$ have the same spectrum
and if 
$\hM_j=\hE_j=
\hPi_j$ 
by substituting $m$ with $j$,  
Eq.~(\ref{eq.POVM1})  is equivalent to  (\ref{eq.probA}).
Thus, in this special case, the measurement turns out to be the projective measurement of $\hA$. 
However, 
in the context of generalized measurement,
we can design not only the projective measurement
but also weak and approximate measurement 
in which we control the measurement strength 
and the backaction caused by the measurement.

\subsection{Measurement of photon polarization}
\label{sec.MPP}
As described in Sec.~\ref{sec.QMPP}, 
the Pauli matrices are the observables of the single photon polarization quibit.
Their measurement outcomes
are $\pm 1$, 
each of which corresponds to one of the two orthogonal polarization states where the single photon is found.
For instance, 
$\sigma_z$ measures the polarization state in $\ket{H}$ or $\ket{V}$,
and $\sigma_x$ measures the polarization state in $\ket{D}$ or $\ket{A}$.

In experimental optics, 
polarization beamsplitters are commonly
used for polarization measurements.
A polarization beamsplitter (PBS) transmits one polarization component (p-component) parallel to the plane of incidence
and reflects the other component (s-component) perpendicular to the plane of incidence.
Hereafter, we take the laboratory coordinates so that p-component is horizontal (H) and  the s-component is vertical (V).  
Thus, a PBS treats the polarization degrees of freedom (H or V) and the path degrees of freedom (1 or 2).
The input--output relation between the field operators is
\begin{eqnarray}
\left(
\begin{array}{c}
\aa_{H1}' \\
\aa_{H2}' \\
\aa_{V1}' \\
\aa_{V2}' \\
\end{array}
\right)
= 
\hU_\mathrm{PBS}
\left(
\begin{array}{c}
\aa_{H1} \\
\aa_{H2} \\
\aa_{V1} \\
\aa_{V2} \\
\end{array}
\right) ,
\label{eq.PBS1}
\end{eqnarray}
where
\begin{eqnarray}
\hU_\mathrm{PBS}
= 
\left(
\begin{array}{cc}
\hU_H & \bm0\\
\bm0 & U_V  
\end{array}
\right) ,
\quad
\hU_H
= e^{i\phi}
\sigma_0 ,
\quad
\hU_V
= 
\sigma_x .
\label{eq.PBS2}
\end{eqnarray}
Here, $\phi$ is the phase difference between the transmitted and reflected components.
Since $\phi$ gives no effect or can be compensated in our experiments,
we set $\phi=0$ in the following.  
When a single photon in these $2\times2$ dimensional modes is concerned,
we can treat the polarization degree of freedom as one qubit having eigenstates $\ket{H}$ and $\ket{V}$,
and the path degree of freedom as the other qubit having eigenstates $\ket{{+1}}$ and $\ket{{-1}}$ for paths 1 and 2, respectively.
The general expression of the two qubit state is
\begin{eqnarray}
 \ket{\Psi} 
 &=& c_{H1} (\ket{H}\otimes \ket{{+1}})  + c_{H2} (\ket{H}\otimes \ket{{-1}}) 
 \nonumber \\
 &+& c_{V1} (\ket{V}\otimes \ket{{+1}})  + c_{V2} (\ket{V}\otimes \ket{{-1}}) . 
\end{eqnarray}
For these bases,
the transform operator of the PBS is the same as $\hU_\mathrm{PBS}$ in (\ref{eq.PBS2}).  
Indeed, it acs as a controlled NOT (CNOT) operation between the two qubits,
with the polarization as control and the path as target:
\begin{eqnarray}
 \hU_\mathrm{PBS}
 = \ketbra{H}{H}\otimes \sigma_0 
 + \ketbra{V}{V}\otimes \sigma_x .
\end{eqnarray}
We can use the PBS for a polarization measurement device using the path quibit as a probe.
We assume that a photon having the polarization state $\ket{\psi}$ initially comes from the path 1, i.e., the initial path qubit is $\ket{{+1}}$,
and the measurement is done by detecting the photon in one of the two output paths after the PBS.
The photon's initial state 
$
\ket{\Psi} = \ket{\psi}\otimes\ket{{+1}}
$ 
is transformed after passing through the PBS as
\begin{eqnarray}
\ket{\Psi'}
 =  \hU_\mathrm{PBS}(\ket{\psi}\otimes\ket{{+1}})
 = \hM_+\ket{\psi}\otimes \ket{{+1}} 
 + \hM_- \ket{\psi}\otimes \ket {{-1}} ,
\end{eqnarray}
where the measurement operators 
$\hM_+$ and $\hM_-$
are given by
\begin{eqnarray}
 \hM_+= \bra{{+1}} \hU_\mathrm{PBS} \ket{{+1}} = \ketbra{H}{H}, \quad 
 \hM_-= \bra{{-1}} \hU_\mathrm{PBS} \ket{{+1}} = \ketbra{V}{V} .
\end{eqnarray}
Thus the measurement is a projective measurement of the photon polarization.

\begin{figure}[t!]
\centerline{
\includegraphics[width=60mm]{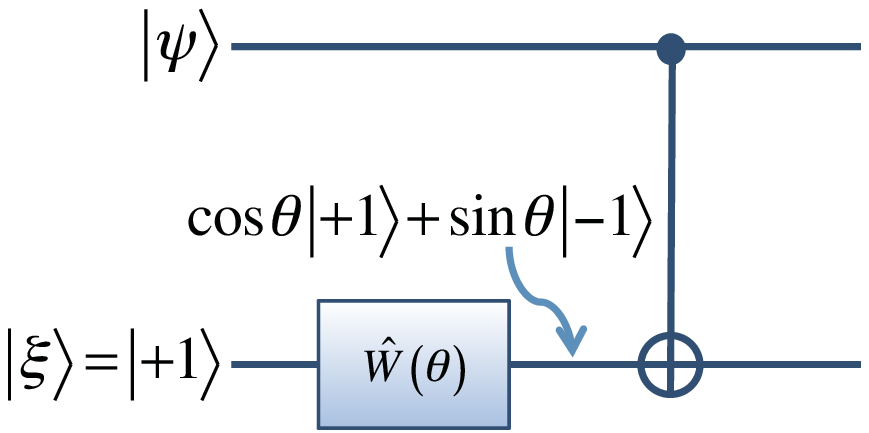}
}
\caption{
A model apparatus for  generalized measurement on a qubit \cite{Lund:2010cn},
where 
$\hW(\theta) \equiv \left( \begin{array}{cc}\cos\theta & \sin\theta \\ \sin\theta & -\cos\theta\end{array} \right) $.
\label{fig.gqmcircuit}
}
\end{figure}

A model apparatus for generalized measurement on a qubit was proposed by Lund and Wiseman \cite{Lund:2010cn}.
As shown in Fig.~\ref{fig.gqmcircuit}, 
the measurement consists of a single qubit gate 
\begin{eqnarray}
\hW (\theta) = 
\left(
\begin{array}{cc}
 \cos\theta & \sin\theta \\
\sin\theta & -\cos\theta 
\end{array}
\right) 
\label{eq.GMP1} 
\end{eqnarray}
on the probe qubit 
(initial state:  $\ket{\xi}=\ket{{+1}}$)
and a succeeding CNOT gate
between the signal qubit $\ket{\psi}$ as control and the probe qubit as target.
It can be shown that the measurement operators of this model apparatus are
\begin{eqnarray}
 \hM_+ 
 = \cos\theta \ketbra{H}{H} + \sin\theta \ketbra{V}{V} 
 = \frac{1}{\sqrt2}\left(\alpha \hI + \beta \sigma_z \right) , 
\label{eq.GMP2a} 
 \\
 \hM_- 
 = \sin\theta \ketbra{H}{H} + \cos\theta \ketbra{V}{V} 
 = \frac{1}{\sqrt2}\left(\alpha \hI - \beta \sigma_z \right) , 
\label{eq.GMP2b} 
\end{eqnarray}
where
$\alpha=\cos(\pi/4-\theta)$ and $\beta=\sin(\pi/4-\theta)$. 
The corresponding POVM elements are
\begin{eqnarray}
 \hE_+ = \cos^2\theta \ketbra{H}{H} + \sin^2\theta \ketbra{V}{V}  = \frac12 \left(\hI + \cos 2\theta\,\sigma_z \right), 
 \\
 \hE_- = \sin^2\theta \ketbra{H}{H} + \cos^2\theta \ketbra{V}{V} = \frac12 \left(\hI - \cos 2\theta\,\sigma_z \right).
\label{eq.GMP3} 
\end{eqnarray}
The measurement is a projective measurement for $\theta=0$,
since in this case $\hM_+$ and $\hM_-$ are the projectors to $\ket{H}$ and $\ket{V}$, respectively.
On the other hand, the measurement is the null measurement for $\theta=\pi/4$;
in this case the measurement for any state returns either outcome with even probability.    
The measurement strength is characterized by $s= \cos 2\theta $ ($0 \le s \le 1$).
Thus this protocol realizes the
generalized measurement with variable measurement strength.

\begin{figure}[bt]
\centerline{
\includegraphics[width=150mm]{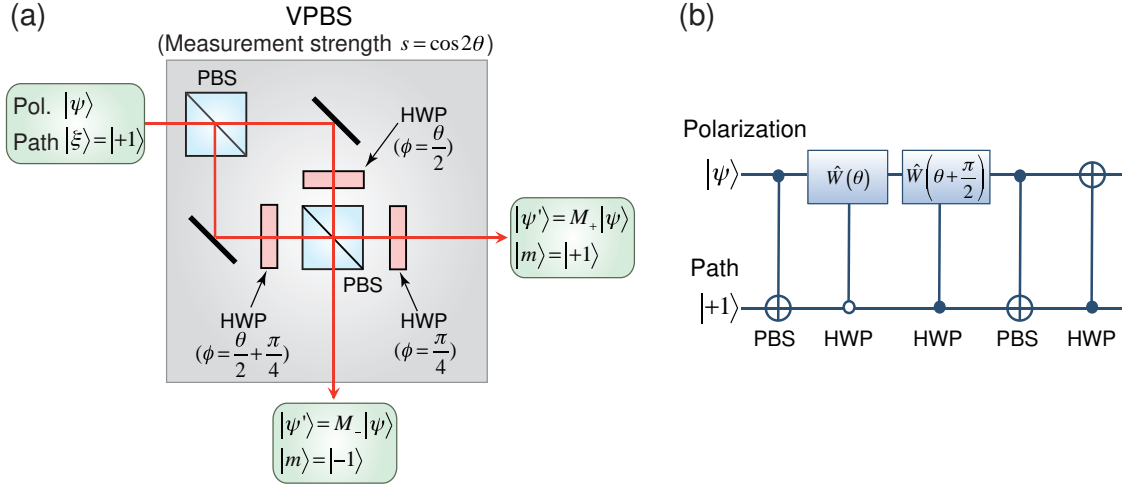}
}
\caption{
(a) Generalized polarization measurement using a variable polarizing beamsplitter (VPBS) \cite{Baek:2008bn,Baek:2013fh}.
PBS and HWP stand for polarization beamsplitters and half-wave plates, respectively.
The signal is a photon polarization qubit $\ket{\psi}$,
and the probe is a path qubit $\ket{\xi}$.
A photon having polarization $\ket{\psi}$ is injected from the path $\ket{\xi}=\ket{{+1}}$,
and exits from either of the output path $\ket{m}=\ket{{+1}}$ or $\ket{{-1}}$
depending on the measurement outcome of the polarization qubit. 
(b) quantum circuit model of the VPBS.
}
\label{fig.vpbs}
\end{figure}

The optical implementation of the generalized measurement described above is shown in Fig.~\ref{fig.vpbs},
which we call a {\it variable polarization beamsplitter} (VPBS) \cite{Baek:2008bn,Baek:2013fh}.
Here, the signal to be measured is the single photon polarization qubit,
and the probe is the path qubit, i.e, the use of path degrees of freedom ($\ket{{+1}}$ and $\ket{{-1}}$) of the photon,
the roles of which are the same as those of a PBS.
The quantum circuit model of this apparatus is shown in Fig.~\ref{fig.vpbs} (b).
Although the circuit is a bit different from that in Fig.~\ref{fig.gqmcircuit},
these are statistically equivalent, i.e., both give the same measurement operators
given in (\ref{eq.GMP2a}) and (\ref{eq.GMP2b}) for the same probe input $\ket{\xi}=\ket{{+1}}$.

\section{Error and disturbance in quantum measurement}

\subsection{Definitions of error and disturbance}

In order to discuss  the accuracy and precision of the measurement, 
we must define the error of the measurement.
In classical science, we assume the existence of true value $x_0$ for the physical quantity to be measured.
The measurement error is often characterized by the root mean square (RMS) distance of the measurement outcomes $x$ and the true value $x_0$
as
\begin{eqnarray}
\Delta x_\mathrm{rms} = \sqrt{\mean{(x-x_0)^2}}
\end{eqnarray}
Note that $\Delta x_\mathrm{rms}$ includes both accuracy and precision, 
as the distance between the true value and the mean value of the outcomes on one hand,
and as the distribution of the measurement outcomes on the other hand.

However, in quantum measurement, 
we cannot assume the true value in general.
Instead, the measurement results are generally probabilistic and their distribution depends on the signal's state.
Also, in general, the object's state is disturbed because of the back-action of the measurement,
resulting in the disturbance on the sequential or joint measurement.  
Because of this property of quantum measurement,  
no commonly agreed definitions of the measurement error and disturbance have been established to date.
Nevertheless, there are a number of proposals on the definitions of error in quantum measurement.
Some of them are defined as state-dependent, 
i.e., the amounts of error and disturbance are dependent on the state of the object to be measured,
and others are defined as state-independent.
Also, some of them are defined based on the RMS of measurement outcomes, 
and others are defined based on information-theoretic quantities.
There are active discussions and debates on this matter \cite{Ozawa:2014jy,Busch:2014dc,Buscemi:2014ew}.
Here, we introduce a state-dependent definition of the measurement error 
given in the general theory of quantum instruments proposed by Ozawa \cite{Ozawa03a}.
 
We consider the generalized measurement model introduced in Sec.~\ref{sec.GM}  (Fig.~\ref{fig.model}).
The signal state $\ket{\psi}$ is subjected to the measurement of an observable $\hA$ 
by an instrument in which $\ket{\psi}$ interacts with the probe state $\ket{\xi}$ through $\hU$.
The measurement outcome is obtained by observing the probe's observable $\hM$ after the interaction.
After the measurement of $\hA$ , the signal is then subjected to the projective measurement of another observable $\hB$.
Using the Heisenberg picture, 
the observables $\hM$ and $\hB$ acting on
the input state
$\ket{\Psi}=\ket{\psi}\otimes\ket{\xi}$
are:
\begin{eqnarray}
\hM_A&=\hU^{-1}(\hI \otimes \hM) \hU ,
\label{ma}\\
\hM_B&=\hU^{-1}(B \otimes \hI) \hU .
\label{mb}
\end{eqnarray}
These are the observables corresponding to what is actually measured by this instrument.
The noise operator $\hN(A)$ and the disturbance operator $\hD(B)$ are defined as 
the difference between the observables that we actually measure and that we want to measure:
\begin{eqnarray}
   \hN(A) &= \hM_A - \hA, 
\label{ed01}\\
   \hD(B) &=  \hM_B - \hB. 
\label{ed02}
\end{eqnarray}
Here and hereafter, we use the abbreviation: 
$\hA \otimes \hI$ as simply $\hA$, and $\hB \otimes \hI$ as $\hB$.    
Then the error $\epsilon(A)$ and the disturbance $\eta(B)$ in the measurement of $\hA$ and $\hB$ are defined as
the RMS of $\hN$ and $\hD$ \cite{Ozawa03a}:
\begin{eqnarray}
 \epsilon(A) &=  \sqrt{ \mean{ \hN(A)^2 } }  , 
\quad
\label{ed1} 
  \eta(B) &=  \sqrt{ \mean{ \hD(B)^2 } }  .
\label{ed2}
\end{eqnarray}
These definitions of error and disturbance were given by Ozawa \cite{Ozawa03a}.
The same or similar definitions 
were proposed and widely used by
Arthurs and Kelly \cite{Arthurs:1965tx},
Arthurs and Goodman \cite{Arthurs:1988fv}, Appleby \cite{Appleby:1998gn,Appleby:1998eb}, 
Hall \cite{Hall:2004cs}, and Branciard \cite{Branciard:2013cb}, etc.
It is important to note that,
if  $\hM_A$ and 
$\hA$
commute, 
Eq.~(\ref{ed1}) corresponds to the classical RMS error \cite{Ozawa:2014jy}．
In this sense, 
Eq.~(\ref{ed1}) is considered to be the generalization of the classical RMS error to the quantum measurement.

\subsection{Evaluation of error and disturbance}
From the definition of the measurement error (\ref{ed1}), we get
\begin{eqnarray}
 \epsilon(A)^2 
& =\mean{ (\hM_A-\hA)^2 } \nonumber\\
& =\mean{\hM_A^2} + \mean{\hA^2} - \mean{\hM_A \hA + \hA \hM_A} ,
\label{eq.ee1}
\end{eqnarray}
and a similar relation for $\eta(B)^2$.
The first two terms of (\ref{eq.ee1}) can be evaluated experimentally or theoretically.
Using (\ref{eq.POVM1}) and (\ref{eq.SD1}),
\begin{eqnarray}
\mean{\hM_A^2} = \sum_m \mu_m^2 \mean{\hE_m} , \quad
\mean{\hA^2} = \sum_j \lambda_j^2 \mean{\hPi_j} .
\label{eq.ee1a}
\end{eqnarray}
For instance, in the case of qubit measurement where $\mu_m=\lambda_j=\pm1$,  
these terms turn out to be unity.
However, the experimental evaluation of the third term, which presents the correlation between $\hM_A$ and $\hA$, is not so 
straightforward.

\begin{figure}[bt]
\centerline{
\includegraphics[width=90mm]{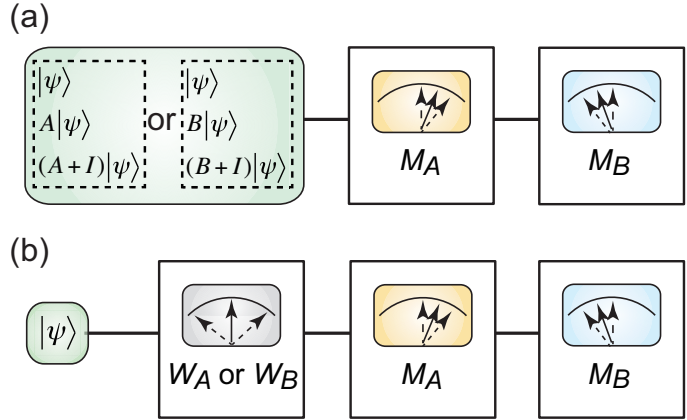}
}
\caption{
Methods to evaluate measurement error and disturbance:
(a) three-state method
and 
(b) weak probe method.
}
\label{fig.meas}
\end{figure}

One method is to transform the third term of (\ref{eq.ee1}) as
\begin{eqnarray}
\mean{\hM_A \hA + \hA \hM_A}
= \mean{(\hI+\hA)\hM_A (\hI+\hA)} -  \mean{\hA \hM_A \hA} - \mean{\hM_A}.
\label{eq.ee2}
\end{eqnarray}
In this form, 
the first term of the right hand side is the expected value of $\hM_A$ for the signal state $(\hA+\hI)\ket{\psi}$.
Also, the second and the third terms are those of the states $\hA\ket{\psi}$ and $\ket{\psi}$, respectively.
Thus,
if these three states are prepared, one can evaluate the experimental error
by (\ref{eq.ee1}) and (\ref{eq.ee2}),
as illustrated in Fig.~\ref{fig.meas}\,(a).
This procedure, called as the three-state method, is given by Ozawa \cite{Ozawa04a}.
Similarly, we show that one may use the relation
\begin{eqnarray}
2 \mean{\hM_A \hA + \hA \hM_A} 
= \mean{(\hI+\hA)\hM_A (\hI+\hA)} - \mean{(\hI-\hA)\hM_A (\hI-\hA)} ,
\label{eq.ee3}
\end{eqnarray}
preparing  $(\hI+\hA)\ket{\psi}$ and $(\hI-\hA)\ket{\psi}$ as the signal states (two-state method).
In the case of qubit measurement, 
$\hA=\sigma_z$ for instance,  $\sigma_z$ presents the rotation on the Bloch sphere,
$(\hI\pm\sigma_z)/2$
are the projectors to the two eigenstates of $\sigma_z$. 
Thus, it is not difficult to prepare these states in experiments.
In practice, the three-state method was used to evaluate the error and disturbance in the measurement of
neutron spin \cite{Erhart:2012dy} and photon polarization \cite{Baek:2013fh}.
However, in general cases, it is difficult to implement the operation $\hI\pm\hA$ or even $\hA$ in practical experiments,
and thus the applicability of (\ref{eq.ee2}) or (\ref{eq.ee3}) is not so obvious. 

The other method 
is to use weak values.
Using the POVM elements $\hE_m$ of the measurement $\hM_A$, we obtain
\begin{eqnarray}
\mean{\hM_A \hA + \hA \hM_A}
& = 2\,\R\!\mean{\hM_A \hA} \nonumber\\
& =2 \sum_m \mu_m\, \R\!\mean{\hE_m \hA} \nonumber\\
& =2 \sum_{ j,m} \lambda_j\mu_m\, \R\!\mean{\hE_m \hPi_j} .
\label{eq.ee4}
\end{eqnarray}
In the last expression, $\R\!\mean{\hE_m \hPi_j}\equiv P_{W}(j,m)$ is called
the {\it weak-valued joint probability} \cite{Wiseman:2003fd},
which
is related to the weak value \cite{Aharonov:1988fk,Steinberg:1995gf}:
\begin{eqnarray}
 \R\mean{\hA}_{W} &=  
 \frac{\R\!\mean{\hE_m \hA }}{\mean{ \hE_m}}
 =  \sum_j \lambda_j \frac{P_{W}(j,m)}{\mean{ \hE_m}} .
\label{eq.ee5}
\end{eqnarray}
Using the relations
(\ref{eq.ee1}), (\ref{eq.ee1a}), (\ref{eq.ee4}), and 
\begin{eqnarray}
 \mean{\hE_m} 
 =  \sum_j  P_{W}(j,m) ,
 \quad
\mean{\hPi_j} 
 =  \sum_m  P_{W}(j,m) ,
\label{eq.ee6}
\end{eqnarray}
one finds \cite{Lund:2010cn,Ozawa:2005dp} 
\begin{eqnarray}
 \epsilon(A)^2 
& = \sum_{j, m} (\mu_m - \lambda_j)^2 P_{W}(j,m) .
\label{eq.ee7}
\end{eqnarray}
Thus, we can evaluate the measurement error if we know every weak-valued joint probability 
$P_{W}(j,m)$.
Lund and Wiseman \cite{Lund:2010cn} derived (\ref{eq.ee7}) 
and presented a practical example using the generalized qubit measurement as described in Sec.~\ref{sec.MPP}.
However, for general $\hA$, 
the experimental evaluation of every weak-valued joint probability is usually impractical.

In what follows, we show a more practical and general procedure to evaluate the measurement error and disturbance using weak measurement or weak probe
(weak probe method, illustrated in Fig.~\ref{fig.meas}\,(b)). 
In order to evaluate $\R\!\mean{\hE_m \hA }$, 
one may use 
a qubit 
as a probe
that interacts with the signal via
\begin{eqnarray}
\hU = \exp ( ig \hA\otimes \sigma_y ) ,
\label{eq.ee8}
\end{eqnarray}
where $g\, (\ge 0)$ is the coupling strength.
The measurement is done by detecting the probe state in either of the eigenstates of $\sigma_z$, i.e., $\ket{{+1}}$ or $\ket{{-1}}$.
Assuming the initial probe state is $(\ket{{+1}}+\ket{{-1}})/\sqrt2$,
the corresponding measurement operators are
\begin{eqnarray}
\hW_\pm = \frac{1}{\sqrt2}\left[ \cos(g\hA) \pm \sin(g\hA) \right] .
\label{eq.ee9}
\end{eqnarray}
When the coupling is sufficiently weak, i.e, $g\mean{\hA} \ll 1$,  Eq.~(\ref{eq.ee9}) can be approximated as
\begin{eqnarray}
\hW_\pm \simeq \frac{1}{\sqrt2} ( \hI \pm g\hA ) .
\label{eq.ee10}
\end{eqnarray}
After this weak measurement, the signal is subject to the main measurement presented by the POVM elements $\hE_m$.
The joint probability $P(w, m)$ obtaining the outcomes of the weak probe in $w=\pm1$ and the main measurement in $m$ is given by
\begin{eqnarray}
P(w, m) 
&=\mean{ \hW_w^\dag \hE_m \hW_w} 
\nonumber \\
&\simeq \frac12 \mean{ (\hI \pm g\hA ) \hE_m (\hI \pm g\hA ) } 
\nonumber \\
&= \frac12 \mean{\hE_m } \pm \frac{g}{2} \mean{\hE_m\hA + \hA\hE_m} + \frac{g^2}{2} \mean{\hA\hE_m\hA}  .
\label{eq.ee11}
\end{eqnarray}
Thus
\begin{eqnarray}
\sum_{w} w P(w, m) 
\simeq  g \mean{\hE_m\hA + \hA\hE_m} 
= 2 g \R\!\mean{\hE_m\hA}  .
\label{eq.ee12}
\end{eqnarray}
Using (\ref{eq.ee1}), (\ref{eq.ee4}) and (\ref{eq.ee12}),
$\R\!\mean{\hE_m \hA }$ and thus the measurement error 
can be evaluated
by measuring the probability $P(w,p)$
within the weak coupling limit:
\begin{eqnarray}
\epsilon(A)^2
\simeq \mean{\hM_A^2} + \mean{\hA^2} 
- \frac{1}{g}\sum_{w,m} w \mu_m P(w,m). 
\label{eq.ee13}
\end{eqnarray}

When the signal observable is also a qubit, e.g., $\hA = \sigma_z$ 
(more generally when $\hA^2=\hI$), 
(\ref{eq.ee9}) reduces to
\begin{eqnarray}
\hW_\pm 
= \frac{1}{\sqrt2} ( \cos g\, \hI \pm \sin g\, \sigma_z ) 
= \frac{1}{\sqrt2} (\alpha \hI \pm \beta \sigma_z ),
\label{eq.ee14}
\end{eqnarray}
where we put $\alpha=\cos g$ and $\beta=\sin g$. 
Thus, in this case, $\hW_\pm$ 
is equivalent to the measurement operators (\ref{eq.GMP2a}) and  (\ref{eq.GMP2b})
for the generalized qubit measurement described in Sec.~\ref{sec.MPP}.
We get
\begin{eqnarray}
P(w, m) 
&=\mean{ \hW_w \hE_m \hW_w} 
\nonumber \\
&= \frac12 \mean{ (\alpha \hI \pm \beta\sigma_z ) \hE_m (\alpha \hI \pm \beta\sigma_z ) } 
\nonumber \\
&= \frac{\alpha^2}{2}\mean{\hE_m } \pm \frac{\alpha\beta}{2} \mean{\hE_m\sigma_z + \sigma_z\hE_m} 
   + \frac{\beta^2}{2}\mean{\sigma_z\hE_m\sigma_z },
\label{eq.ee15}
\end{eqnarray}
%
%
\begin{eqnarray}
\sum_{w} w P(w, m) 
= 2 \alpha\beta\, \R\!\mean{\hE_m\sigma_z}  .
\label{eq.ee16}
\end{eqnarray}
Thus we obtain
\begin{eqnarray}
\epsilon(A)^2
&= 2 - \frac{1}{\alpha\beta}\sum_{w,m} w \mu_m P(w,m) 
 \nonumber\\
&=  2- \frac{2}{\sin 2g}\sum_{w,m} w \mu_m P(w,m) .
\label{eq.ee17}
\end{eqnarray}
Here, $2\alpha\beta=\sin 2g$ is the measurement strength of the weak probe.
Note that
in the qubit case the coupling strength $g$ is not necessarily weak;
Eqs.~(\ref{eq.ee16})  and (\ref{eq.ee17}) are valid for any $g$.
Thus, we can even use the projective measurement where $\alpha=\beta=1/\sqrt2$.
In this case, 
it is interesting to observe that
(\ref{eq.ee16}) is equivalent to the procedure obtained in
the two-state method (\ref{eq.ee3}).

\section{Uncertainty relations in quantum measurement}
\label{sec.UR}

\subsection{Heisenberg's relation}

In 1927,
Heisenberg considered the relationship between the measurement error and disturbance 
in his famous thought experiment of gamma-ray microscope \cite{Heisenberg:1927gz}.
His relation is written as
\begin{eqnarray}
  \epsilon(x) \eta(p)  \ge \frac{\hbar}{2},
\label{eq.Heisenberg1}
\end{eqnarray}  
where $\epsilon(x)$ is the measurement error of the position $x$, 
and $\eta(p)$ the disturbance in the momentum $p$.
He obtained Eq.~(\ref{eq.Heisenberg1}), under some assumptions \cite{Ozawa:2015vy}, from the relation proven by Kennard \cite{Kennard:1927hl}
\begin{eqnarray}
  \sigma(x) \sigma(p)  \ge \frac{\hbar}{2}
\label{eq.Kennard}
\end{eqnarray}  
between the standard deviations (e.g., $\sigma(x)=\sqrt{\mean{x^2}-\mean{x}^2}$) of $x$ and $p$.
Later, Arthurs and Kelly \cite{Arthurs:1965tx} quantified and confirmed the relation (\ref{eq.Heisenberg1})
in the case where the measurement of $x$ and $p$ are both unbiased,
i.e., the mean value of the measurement results are the same as that of the corresponding observable for every state.
The generalization of Eq.~(\ref{eq.Kennard}) is
\begin{eqnarray}
  \sigma(A) \sigma(B)  \ge C,
\label{eq.Robertson}
\end{eqnarray}
where 
\begin{eqnarray}
%
  C \equiv \frac12  \left| \mean{ [ \hA, \hB  ] } \right| .
\label{eq.URC}  
\end{eqnarray}
Eq.~(\ref{eq.Robertson}) was proven by Robertson \cite{Robertson:1929kt} and is referred to as Robertson's relation.
The relation between the standard deviations as in (\ref{eq.Kennard}) and (\ref{eq.Robertson})  
is sometimes referred to as the uncertainty relation in state preparation, or preparation uncertainty relation.
On the other hand, the relation between the measurement error and disturbance as in (\ref{eq.Heisenberg1})
is referred to as the uncertainty relation in joint measurement, error-disturbance relation (EDR), or measurement-disturbance relation (MDR).
The generalization of 
Heisenberg's relation (\ref{eq.Heisenberg1}) corresponding to (\ref{eq.Robertson}) 
is \cite{Arthurs:1988fv,Ozawa:1991km,Ishikawa:1991jp} 
\begin{eqnarray}
  \epsilon(A) \eta(B)  \ge C .
\label{eq.Heisenberg2}
\end{eqnarray}
%
Eq.~(\ref{eq.Heisenberg2}) 
is sometimes referred to also as 
Heisenberg's relation.
It is known that the relation (\ref{eq.Heisenberg2}) is valid  
under the assumption that the measurements are both unbiased for $\hA$ and $\hB$ \cite{Arthurs:1988fv,Ozawa:1991km,Ishikawa:1991jp}. 
However, it is noteworthy that the relation  (\ref{eq.Heisenberg1}) or (\ref{eq.Heisenberg2}) 
may be violated 
if this assumption does not hold.

\subsection{Ozawa's and Branciard's relations}
In 2003, 
using the definitions of error and disturbance in 
(\ref{ed1})  
and Robertson's relation (\ref{eq.Robertson}),
Ozawa derived the universally valid relation \cite{Ozawa03a}
\begin{eqnarray}
  \epsilon(A) \eta(B) +  \frac12 \left| \mean{[\hN(A),\hB]} + \mean{[\hA, \hD(B)]} \right| \ge C .
\label{eq.Ozawa0}
\end{eqnarray}
When $\hN(A)$ and $\hD(B)$ both give constant values regardless of the state, i.e.,
\begin{eqnarray}
\mean{\hN(A)}=\mean{\hM_A}-\mean{\hA}=a ,
\label{HeisenbergNA1}\\
\mean{\hD(B)}=\mean{\hM_B}-\mean{\hB}=b ,
\label{HeisenbergNB1}
\end{eqnarray}
the second and the third terms of (\ref{eq.Ozawa0}) vanish.
Thus, in this case, Eq.~(\ref{eq.Ozawa0}) is reduced to Heisenberg's relation (\ref{eq.Heisenberg2}). 
If  $M_A-a$ and $M_B-b$ are redefined as $M_A$ and $M_B$, respectively,
we get
\begin{eqnarray}
\mean{\hM_A}-\mean{\hA}=0 ,
\label{HeisenbergNA}\\
\mean{\hM_B}-\mean{\hB}=0 .
\label{HeisenbergNB}
\end{eqnarray}
These are nothing else than the unbiased conditions of the measurements of $\hM_A$ and $\hM_B$.
Thus, Eq.~(\ref{eq.Ozawa0}) is regarded as the generalization of 
Heisenberg's relation (\ref{eq.Heisenberg2}),
specifying the forming condition of 
Heisenberg's relation.  
From (\ref{eq.Ozawa0}), 
Ozawa derived 
another universally valid relation \cite{Ozawa03a}
\begin{eqnarray}
  \epsilon(A) \eta(B) +  \epsilon(A) \sigma(B) + \sigma(A) \eta(B) \ge C 
\label{eq.Ozawa}
\end{eqnarray}
between the measurement error, disturbance and standard deviations.
In the left hand sides, (\ref{eq.Ozawa0}) and (\ref{eq.Ozawa}) contain additional terms 
that are absent from 
Heisenberg's relation (\ref{eq.Heisenberg2}),
while the right hand sides are the same.
Thus, the term $\epsilon(A) \eta(B)$ itself may be smaller than the right hand side,
suggesting that it is possible to violate 
Heisenberg's relation (\ref{eq.Heisenberg2}).
Given the definitions of measurement error and disturbance in (\ref{ed1}),
Eqs.~(\ref{eq.Ozawa0}) and (\ref{eq.Ozawa}) are more general relations 
including 
Heisenberg's relation (\ref{eq.Heisenberg2}) as a special case.

Although the relation (\ref{eq.Ozawa}) is universally valid,
it is not tight in general;
the left hand side is always greater than the right hand side. 
It may be tight only for cases where $\epsilon(A)=0$ or $\eta(B)=0$ \cite{Ozawa03a}.
In 2013, based on the definitions of measurement error and disturbance (\ref{ed1}), 
Branciard  derived the stronger relation \cite{Branciard:2013cb}
\begin{eqnarray}
   \epsilon(A)^2\sigma(B)^2 + \sigma(A)^2\eta(B)^2 
  + 2\epsilon(A) \eta(B)\sqrt{\sigma(A)^2\sigma(B)^2-C^2}  \ge C^2 .
\label{eq.Branciard1}
\end{eqnarray}
Branciard's relation (\ref{eq.Branciard1}) is proven to be universally-valid 
and tight, i.e., there exist cases where the left and right hand sides are equal, 
for general joint measurements of $\hA$ and $\hB$.
A simpler expression can be derived from (\ref{eq.Branciard1}):
\begin{eqnarray}
 \epsilon(A) \sigma(B) + \sigma(A) \eta(B) \ge C, 
\label{eq.Branciard1a}
\end{eqnarray}
which is just the second and third terms of Ozawa's relation (\ref{eq.Ozawa}).
Hence, Branciard's relation (\ref{eq.Branciard1}) is the stronger relation that includes Ozawa's relation (\ref{eq.Ozawa}).
In addition, Branciard derived the even stronger relation 
\begin{eqnarray}
\tilde\epsilon (A)^2  +  \tilde\eta (B)^2
 + 2  \tilde\epsilon (A) \tilde\eta (B) \sqrt{ 1-C^2}  \ge C^2 ,
\label{eq.Branciard2}
\end{eqnarray}
where 
$\tilde\epsilon=\epsilon\sqrt{1-\epsilon^2/4}$
and
$\tilde\eta=\eta\sqrt{1-\eta^2/4}$.
The relation (\ref{eq.Branciard2}) is valid when, as in the case of our photon polarization measurement, 
the spectra of $\hA$, $\hB$ and $\hM$ are all $\pm 1$, and $\mean{A}=\mean{B}=0$ (hence $\sigma(A)=\sigma(B)=1$).

Branciard's relations (\ref{eq.Branciard1}) and (\ref{eq.Branciard2}) are known to be tight for pure signal states.
These relations can be modified to more general relations 
that are tight even for mixed signal states \cite{Ozawa:2014wj}.
To do so,  in (\ref{eq.Branciard1}) and (\ref{eq.Branciard2}),  
we just replace
$C$ defined in (\ref{eq.URC}) with $D$:
\begin{eqnarray}
C \to 
 D = \frac12 \mathrm{Tr} \left|\, \sqrt{\rho}\, [\hA,\hB] \sqrt{\rho}\, \right| ,
\label{eq.OzawaBranciard3}
\end{eqnarray}
where $\rho$ is the density operator of the input state
and $|\hX|$ for an operator $\hX$ is 
a non-negative Hermitian operator 
given by the polar decomposition: $\hX=\hU|\hX|$.

\subsection{Other relations}
As mentioned above,
there are active discussions and debates on the definitions of error and disturbance in quantum measurement.  
Correspondingly, there are a number of proposals for the measurement uncertainty relations 
based on different definitions of error and disturbance.
Busch {\it et al.}\ \cite{Busch:2014dc,Busch:2013dt,Busch:2014ej} proposed a definition of  measurement error  
based on the RMS distance of the distributions between the original ($\hA$) and the measurement ($\hM_A$) observables.
The state-independent error is defined by taking the supremum of the RMS distance with respect to all the input states.
They derived the EDR for the qubit case \cite{Busch:2014ej} 
\begin{eqnarray}
\Delta (A)^2  +  \Delta (B)^2
   \ge 
\sqrt2 \left( \|\bm{a}-\bm{b}\|+\|\bm{a}+\bm{b}\| -2 \right) .
\label{eq.BuschQubit1}
\end{eqnarray}
where $\Delta(A)$ and $\Delta(B)$ are the error and disturbance, under their definition, in the measurements of $\hA$ and  $\hB$, respectively.
In the right hand side, 
$\bm{a}$ and $\bm{b}$ are unit vectors on the Bloch sphere; 
the projection operators $\hPi_{\pm}^{(A)}$ and $\hPi_{\pm}^{(B)}$ for the observables $\hA$ and $\hB$ 
are expressed in terms of $\bm{a}$  and $\bm{b}$ as
\begin{eqnarray}
\hPi_{\pm}^{(A)} =\frac12 (I\pm\bm{a}\cdot\bm{\sigma}), 
\quad
\hPi_{\pm}^{(B)} =\frac12 (I\pm\bm{b}\cdot\bm{\sigma}) ,
\end{eqnarray}
where $\bm{\sigma} = (\sigma_x, \sigma_y, \sigma_z)$ in (\ref{eq.PauliOperators}).
They also showed that in the qubit case $\Delta(A)$ and $\Delta(B)$ coincide with $\epsilon (A)$ and  $\eta (B)$, respectively, 
and thus 
\begin{eqnarray}
\epsilon (A)^2  +  \eta (B)^2
   \ge 
\sqrt2 \left( \|\bm{a}-\bm{b}\|+\|\bm{a}+\bm{b}\| -2 \right) .
\label{eq.BuschQubit2}
\end{eqnarray}
When $\bm{a} \perp \bm{b}$, 
the right hand side of  (\ref{eq.BuschQubit2}) is maximized as
\begin{eqnarray}
\epsilon (A)^2  +  \eta (B)^2
   \ge 2 \left( 2-\sqrt2 \right) .
\label{eq.BuschQubit3}
\end{eqnarray}

Information-theoretic definitions of error, disturbance and EDR in quantum measurements
were recently considered by
Hofmann \cite{Hofmann:2003cg}, 
Watanabe {\it et al.}\ \cite{Watanabe:2011dy}, 
Buscemi {\it et al.}\ \cite{Buscemi:2014ew},
Coles and Furrer \cite{Coles:2015cm}, 
and
Sulyok {\it et al.}\ \cite{Sulyok:2015ct}.
These are based on the  information-theoretic quantities, e.g., conditional entropy,  
that quantify the uncertainty in the estimation (or retrodiction) of the value of original observable
from the measurement outcome.
Some of them assume that the input state is completely unknown, i.e., it is in a fully mixed state \cite{Buscemi:2014ew,Coles:2015cm};
thus the corresponding definitions of error, disturbance and EDR are state-independent.
Since the definitions
are different,
the translation of these relations to the RMS-based relations is not straightforward.
For instance, Buscemi {\it et al.}\ \cite{Buscemi:2014ew} translated their relation for the qubit case
to the RMS-based relation as
\begin{eqnarray}
\left[ \epsilon (A)^2 +\frac13 \right] 
\left[ \eta (B)^2 +\frac13 \right] 
   \ge 
\left( \frac{4}{\pi e} \right)^2
\simeq 0.219 .
\label{eq.BuscemiQubit1}
\end{eqnarray}
However, this relation is weaker than the tight relation predicted for the maximally mixed states 
defined in (\ref{eq.Branciard2}) and (\ref{eq.OzawaBranciard3})
\cite{Ozawa:2014wj}.
Recently, a tight relation within the framework of the information-theoretic definition was proposed \cite{Sulyok:2015ct}, 
but its translation to the RMS-based relation is not apparent.

\section{Experiments on the measurement error, disturbance, and uncertainty relations}

To date, experimental evaluation of the measurement error and disturbance 
in qubit systems has been reported 
using neutron spin \cite{Erhart:2012dy,Sulyok:2015ct,Sulyok:2013gm} and photon polarization \cite{Baek:2013fh,Rozema:2012bx,Weston:2013fe,Kaneda:2014em,Ringbauer:2014gk}.
Here, we review our experiments \cite{Baek:2013fh,Kaneda:2014em}
in which the measurement error, disturbance and uncertainty relations were examined 
in generalized, strength-variable measurement of a single photon polarization.

\begin{figure}[bt]
\centerline{
\includegraphics[width=95mm]{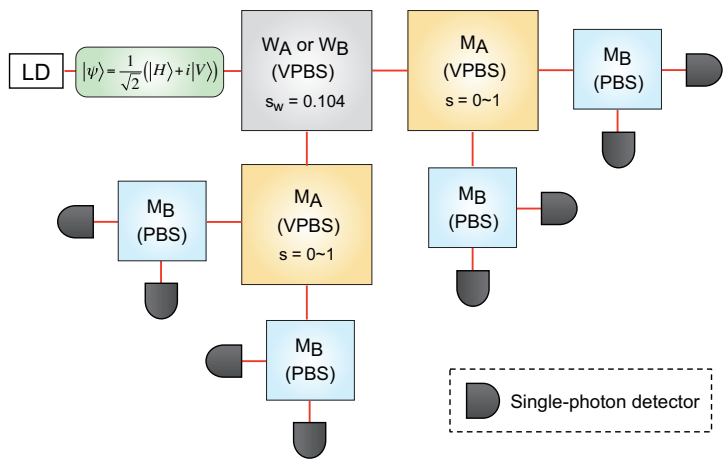}
}
\caption{
Schematic diagram of the experimental setup. 
The photon source is a strongly attenuated laser diode (LD)
and the polarization is set to $\ket{\psi}=\ket{L}=(\ket{H}+i\ket{V})/\sqrt2$.
The weak probe ($\mathrm{W_A}$ or $\mathrm{W_B}$) 
and the main apparatus ($\mathrm{M_A}$) are based on the VPBS depicted in Fig.~\ref{fig.vpbs} 
but in this experiment they are modified to the Sagnac configuration. 
The projective measurement of $\hB$ is implemented by the PBS.
To evaluate the measurement error in $\hA$ (disturbance in $\hB$), 
the weak probe $\mathrm{W_A}$ ($\mathrm{W_B}$) is chosen to probe $\hA$ ($\hB$), 
with a weak measurement strength ($s_w=0.104$).
Then the main apparatus ($\mathrm{M_A}$) measures $\hA=\sigma_z$ (in H or V polarization)
with the variable measurement strength ($s=0 \sim 1$).
Finally, the post measurement apparatus $\mathrm{M_B}$ measures $\hB=\sigma_x$ (in $\pm45^\circ$ polarization).  
Since each apparatus has two output paths as the measurement outcomes, 
the photon is finally detected in either of the $2^3=8$ output path modes,
depending on the measurement outcomes of  
$\mathrm{W_A}$ (or $\mathrm{W_B}$), $\mathrm{M_A}$, and $\mathrm{M_B}$.
Details are shown in Ref.~\cite{Kaneda:2014em}
}
\label{fig.setup}
\end{figure}

The main measurement apparatus is based on the generalized qubit measurement \cite{Lund:2010cn} described in Sec.~\ref{sec.MPP}.
The optical implementation of the measurement apparatus is shown in Fig.~\ref{fig.vpbs},
i.e.,  a VPBS \cite{Baek:2008bn,Baek:2013fh}.
%
Using the VPBS, we want to measure the photon's polarization in the $\ket{H}$ or $\ket{V}$ basis,
i.e., $\hA=\sigma_z$ for the polarization qubit.
The measurement is done by observing the photons in either of the output path $\ket{{+1}}$ or $\ket{{-1}}$, 
i.e., $\hM=\sigma_z$ for the path qubit.
Then, we make the successive measurement on the photons polarization in the $\ket{D}$ or $\ket{A}$ basis as the observable $\hB$,
i.e., $\hB=\sigma_x$ for the polarization quibit.
In this case, 
$C$ in (\ref{eq.URC}) that appears in uncertainty relations is 
$
C= \left| \mean{[\sigma_z,\sigma_x]} \right|/2  = \left| \mean{\sigma_y} \right|  .
$
Thus, the initial state of the signal, polarization qubit, is chosen to be $\ket{L}$ (or $\ket{R}$), 
so that it maximizes $C=1$ in the uncertainty relations to be examined.

Under the condition described above,
the expected values of the noise 
and disturbance 
operators defined in (\ref{ed01}) and (\ref{ed02})  are
\begin{eqnarray}
\mean{\hN(A)}
&= \mean{\hM_A} - \mean{\hA}
= \left( \cos 2\theta-1 \right) \mean{\sigma_z}, 
\label{eq.ena1}
\\
\mean{\hD(B)}
&= \mean{\hM_B} - \mean{\hB}
= \left( \sin 2\theta-1 \right) \mean{\sigma_x},
\label{eq.enb1}
\end{eqnarray}
where $\theta$ is the parameter given in (\ref{eq.GMP1}), defining the measurement strength as
$s=\cos 2\theta$. 
We see that the expected values (\ref{eq.ena1}) and (\ref{eq.enb1}) are dependent on the input signal state and thus 
the forming conditions (\ref{HeisenbergNA1}) and (\ref{HeisenbergNB1}) 
for Heisenberg's relation (\ref{eq.Heisenberg2}) 
are not fulfilled.
The expected measurement error and disturbance defined in (\ref{ed1}) 
are calculated to be \cite{Lund:2010cn,Baek:2013fh,Kaneda:2014em}
\begin{eqnarray}
\epsilon(A) = 2\sin\theta, \quad
\eta(B) = 2\sin\left( \frac{\pi}{4}-\theta \right).
\label{eq.ed}
\end{eqnarray}
%
Thus, 
for this particular measuring apparatus, both the error and the disturbance are independent 
of the input signal state. 
The error $\epsilon(A)$ and the disturbance $\eta(B)$ remain finite 
even when the other goes to zero when $\theta=0$ or $\pi/4$, 
since the error and disturbance are given by RMS difference between $\pm1$-valued observables. 
At this point,  the violation of 
Heisenberg's relation (\ref{eq.Heisenberg2}) is already apparent.

In the experiment \cite{Kaneda:2014em},
we use the weak probe method to evaluate the measurement error and disturbance.
The experimental setup is illustrated in Fig.~\ref{fig.setup}. 
We use the VPBS for the weak probe ($\mathrm{W_A}$ or $\mathrm{W_B}$) and the main apparatus ($\mathrm{M_A}$), and the PBS for the post measurement of $\hB$.
In this experiment, 
we employed the displaced Sagnac configuration
that provides much higher phase stability than the Mach-Zehnder configuration (Fig.~\ref{fig.vpbs})
used in our previous experiment \cite{Baek:2013fh}.
Using this apparatus,
we evaluated the measurement error $\epsilon(A)$ and disturbance $\eta(B)$
by varying the measurement strength $s$=$\cos 2\theta$ of $\mathrm{M_A}$
from thenull measurement $s$=0 to the projective measurement $s$=1.
In the experiment, the measurement strength of the weak probe ($\mathrm{W_A}$ or $\mathrm{W_B}$)
was set to $\cos 2 \theta_w = 0.104$ 
that produced very small disturbance in the initial signal state.
In practice, for the signal state after the weak probe, we expected $C = 0.995$,
which was close to the ideal value $C=1$.

\begin{figure}[bt]
\centerline{
\includegraphics[width=85mm, clip]{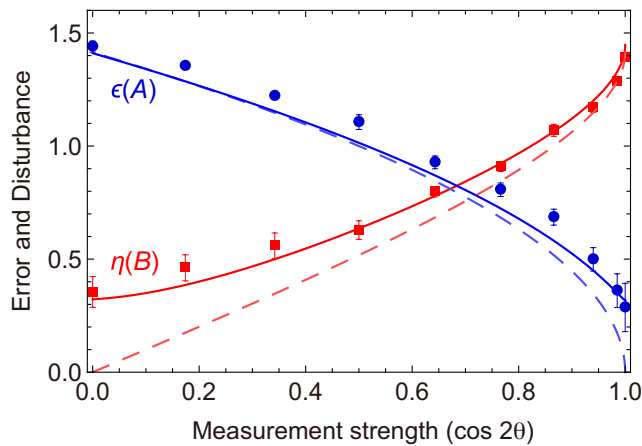}%
}
\caption{Experimental results. 
The error $\epsilon (A)$ (blue circles) and disturbance $\eta (B)$ (red squares) 
are plotted as functions of the measurement strength 
$\cos 2\theta$. 
Dashed curves are the theoretically calculated error and disturbance for perfect implementation of the quantum circuit presented in Fig.~\ref{fig.gqmcircuit} or Fig.~\ref{fig.vpbs}\,(b). 
Solid curves are the theoretical values after the non-ideal extinction ratio of a PBS is taken into account. 
}
\label{fig.results}
\end{figure}

The quantities of $\epsilon (A)$ and $\eta (B)$ thus obtained are shown in Fig.~\ref{fig.results}. 
The dashed curves represent the theoretical calculations of  $\epsilon (A)$ and $\eta (B)$ 
assuming the ideal instrument,
and the solid curves are those in which the imperfect extinction ratio of the PBS is taken into account
(a detailed discussion is given in Refs.~\cite{Lund:2010cn,Baek:2013fh}). 
The experimentally measured error and disturbance present good agreement with the theoretical calculations. 
We clearly see the trade-off relation between the error and disturbance; as the measurement strength increases, $\epsilon(A)$ decreases while $\eta(B)$ increases. 
\begin{figure}[bt]
\centerline{
\includegraphics[width=70mm]{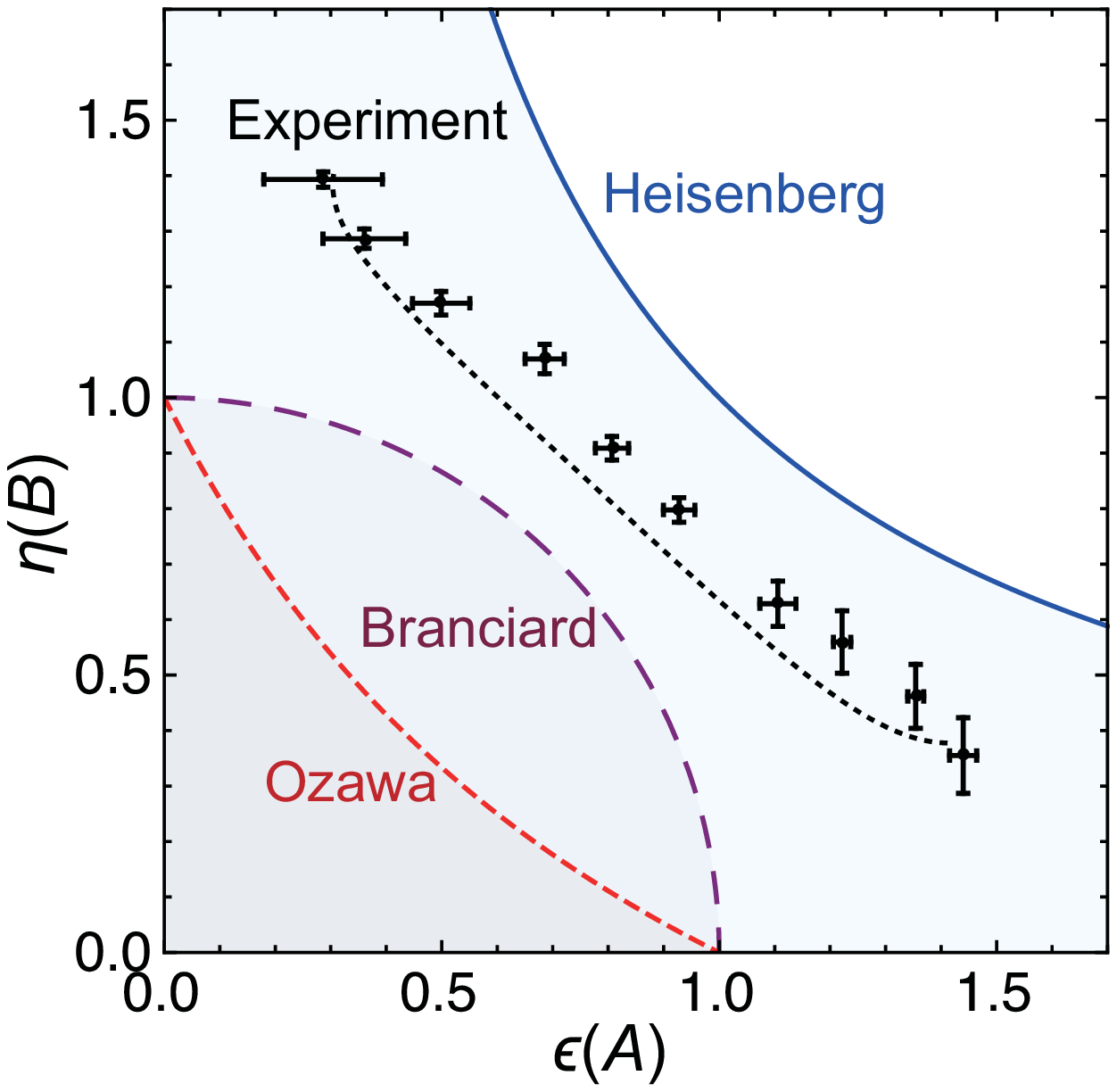}
\hspace{3mm}
\includegraphics[width=70mm]{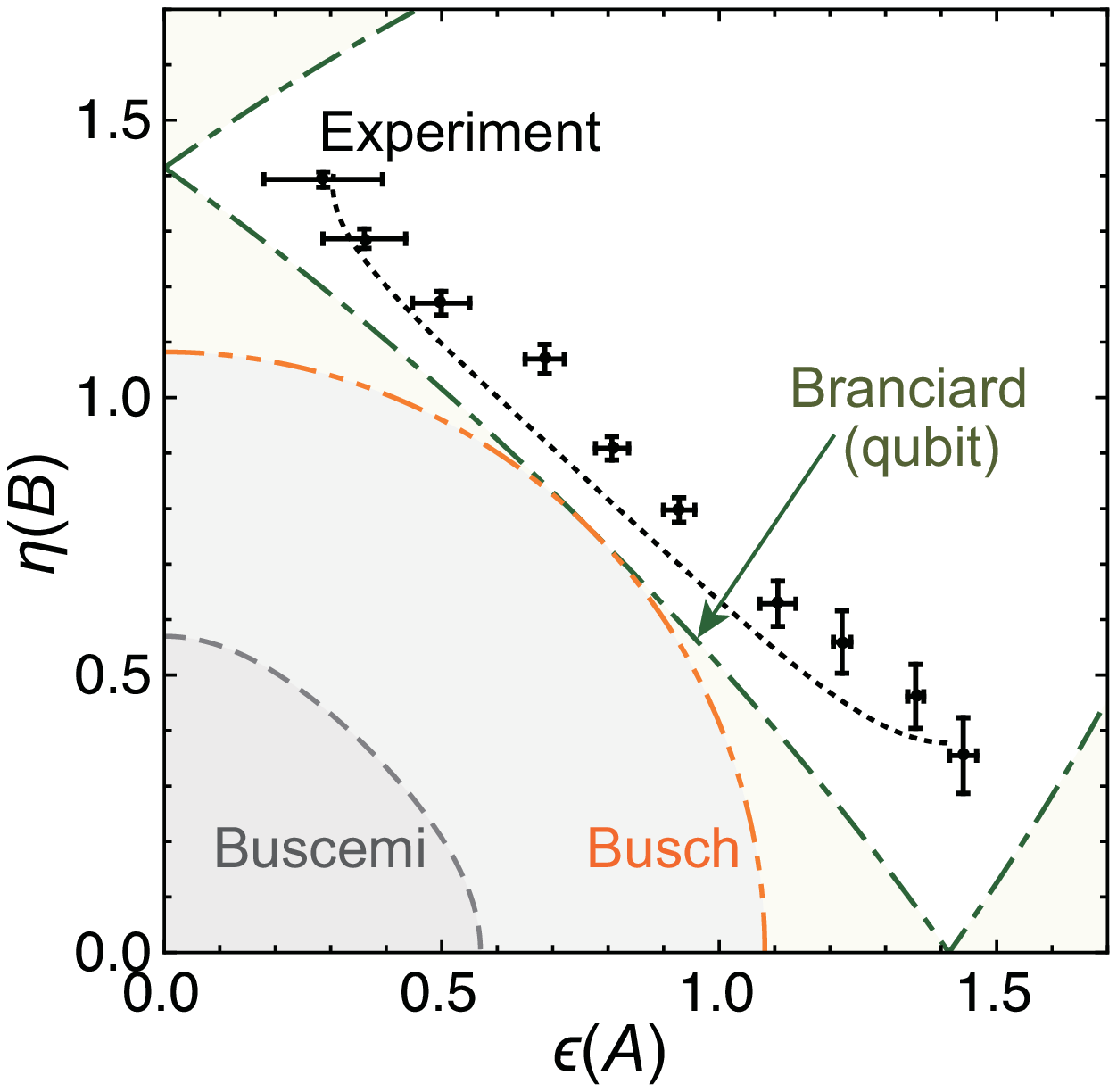}
}
\caption{
Comparison of error-disturbance relations.
Relations for the general case are shown in the left. 
Blue solid curve: Heisenberg's bound in Eq.~(\ref{eq.Heisenberg2}). 
Red (short dashed) curve: Ozawa's bound in (\ref{eq.Ozawa}).
Purple (long dashed) curve: Branciard's bound in (\ref{eq.Branciard1}). 
Relations for the qubit case are shown in the right. 
Green (dot chain) curve: Branciard's bound in (\ref{eq.Branciard2}). 
Orange (two-dot chain) curve: Busch's bound in (\ref{eq.BuschQubit3}). 
Gray (dashed) curve: Buscemi's bound in (\ref{eq.BuscemiQubit1}). 
Black (filled) circles: experimental data shown in Fig.~\ref{fig.results}.
Black (dotted) curve: theoretical prediction for our experiment using imperfect PBSs.
The lower-left side of each bound is the region forbidden by the corresponding EDR.
}
\label{fig.edr}
\end{figure}

In Fig.~\ref{fig.edr}, we plot 
the predicted lower bounds of the EDRs in 
Eqs.~(\ref{eq.Heisenberg2}), (\ref{eq.Ozawa}), 
(\ref{eq.Branciard1}), (\ref{eq.Branciard2}), 
(\ref{eq.BuschQubit3})  and (\ref{eq.BuscemiQubit1}),
together with the experimental data.
Under Heisenberg' EDR the error or disturbance must be infinite when the other goes to zero,  
while other EDRs allow finite error or disturbance even when the other is zero.
We see that the experimental data clearly violate Heisenberg's EDR, 
yet satisfy other recently proposed EDRs.
In particular, our experimental data were close to Branciard's bound (dot chain curve) given in Eq.~(\ref{eq.Branciard2}),   
which could be saturated by ideal experiments.


\section{Conclusions}
We experimentally 
implemented the generalized, strength-variable measurement of photon polarization,
and evaluated the measurement error and disturbance 
making use of weak measurement with minimum disturbance that keeps the initial signal state practically unchanged. 
Our measurement results were compared with various EDRs predicted thus far,
demonstrating
the violation of Heisenberg's EDR
and the validity of Ozawa's and other recently proposed EDRs.

Measurement error, disturbance, and the uncertainty relations are fundamentals 
to our observation of the quantum world.
Recent progress in the quantum theory of measurement has revealed the new aspects of these issues,
providing more precise and fundamental understanding of what we can take from nature through measurements.
Although the experiments thus far carried out on these issues are still limited to qubit systems 
such as a neutron spin or a photon polarization, 
experimental investigation extending to other systems will be 
essential
not only for understanding fundamentals of physical measurement 
but also for developing novel quantum information and communication protocols. 

\section*{Acknowledgement}
The experiments \cite{Baek:2013fh,Kaneda:2014em} described in this article were carried out with the author's former collaborators,
S.-Y. Baek and F. Kaneda.
The author is deeply grateful to M. Ozawa for his theoretical support.
The author also thanks to C. Branciard for valuable discussions.
This work was supported by MIC SCOPE No. 121806010.

\vspace{5mm}

\ifbibtex

\begin{thebibliography}{10}
\bibitem{PauliMaticesNote}
The labeling of Pauli matrices in Eq.\,(\ref{eq.PauliOperators}) is chosen
so that $\sigma_1$, $\sigma_2$ and $\sigma_3$ represent 
the corresponding Stokes operators  $\s_1$, $\s_2$ and $\s_3$
defined in Eqs.~(\ref{ho16-12-1})--(\ref{ho16-12-3}), 
respectively.
If the alternative definition 
was used,
we would obtain $\sigma_1=\sigma_x$, $\sigma_2=\sigma_y$ and $\sigma_3=\sigma_z$.
\end{thebibliography}
\else

\providecommand{\newblock}{}

\fi

\end{document}